\documentclass[useAMS,usenatbib]{mn2e}
\usepackage{graphicx}
\usepackage{deluxetable} 
\def\simlt{\lower.5ex\hbox{\simlt}}
\def\gtsima{$\; \buildrel > \over \sim \;$}
\def\simgt{\lower.5ex\hbox{\gtsima}}

\def\aj{\rm AJ}
\def\apj{\rm ApJ}
\def\aap{\rm A\&A}
\def\mnras{\rm MNRAS}

\title[Multiple populations in NGC~2419]{Evidence for multiple populations in the massive globular cluster NGC~2419 from deep $u$VI LBT photometry\thanks{Based on data acquired using the Large Binocular Telescope (LBT). The LBT is an international collaboration among institutions in the United States, Italy, and Germany. LBT Corporation partners are The University of Arizona on behalf of the Arizona university system; Istituto Nazionale di Astrofisica, Italy; LBT Beteiligungsgesellschaft, Germany, representing the Max-Planck Society, the Astrophysical Institute Potsdam, and Heidelberg University; The Ohio State University; and The Research Corporation, on behalf of The University of Notre Dame, University of Minnesota and University of Virginia.}}
\author[G. Beccari et al.]{G. Beccari$^{1}$\thanks{E-mail: gbeccari@eso.org}, M. Bellazzini$^{2}$, C. Lardo$^{3}$, A. Bragaglia$^{2}$, E. Carretta$^{2}$,  \and
E. Dalessandro$^{3}$, A. Mucciarelli$^{3}$, E. Pancino$^{2}$
\\
$^{1}$European Southern Observatory, Karl-Schwarzschild-Str. 2, 85748 Garching bei Munchen, 
Germany\\
$^{2}$INAF-Osservatorio Astronomico di Bologna, Via Ranzani 1, 40127, Bologna, Italy.\\
$^{3}$Dip. di Astronomia - Univ. di Bologna, Via Ranzani 1, 40127, Bologna, Italy.\\
}
\begin{document}

\date{}

\pagerange{\pageref{firstpage}--\pageref{lastpage}} \pubyear{2011}

\maketitle

\label{firstpage}

\begin{abstract}
We present accurate wide-field $u$VI photometry of the remote and massive Galactic globular cluster NGC~2419, aimed at searching for the $u$-V color spread along the Red Giant Branch (RGB) that is generally interpreted as the photometric signature of the presence of multiple populations in globular clusters. Focusing on the RGB stars in the magnitude range $19.8\le$V$\le22.0$, we find that (a) the $u$-V, $u$-I and the V-I spreads about the RGB ridge line are significantly larger than that expected from observational errors, accurately quantified by means of extensive artificial stars experiments, (b) the intrinsic color spread in $u$-V and $u$-I are larger than in V-I, (c) the stars lying to the blue of the RGB ridge line in $u$-V and $u$-I are significantly more concentrated toward the cluster center than those lying to the red of that line. All the above observational facts can be interpreted in a scenario where a sizable fraction of cluster stars belong to a second generation heavily enriched in Helium. Finally we find 
that bright RGB stars ($17.5<$V$<19.0$) having [Mg/Fe]$<0.0$ lie preferentially on the red side of the cluster RGB, while those having [Mg/Fe]$>0.0$ lie preferentially on the blue side.
\end{abstract}


\section{Introduction}

It is now generally recognised that Galactic Globular Clusters (GCs) are not hosting a pure single-age single-chemical-composition stellar population, but were in fact the site of two (or more) bursts of star formation, accompanied by chemical evolution, during the first $\sim 10^8$ years of their lifetimes as stellar systems \citep[see][G12 hereafter, for a recent thorough review and references]{grat12}. 
This postulation is motivated by the discovery that many GCs host stellar populations with differences in their chemical composition. 

High-resolution spectroscopy of large samples of Red Giant Branch (RGB) stars in many GCs have revealed the presence of significant (and correlated) spreads in the chemical abundance of some light elements (primarily O, Na, Mg, Al) among the stars of each cluster, in spite of a virtually perfect homogeneity in iron abundance \citep[see, e.g.][references and discussion therein]{euge6752tut}. The finding of the same abundance pattern among GC Main Sequence (MS) and Sub Giant Branch (SGB) stars indicates that it is not attributable to evolutionary effects
but to chemical pollution of pristine gas from which cluster stars originally formed \citep[see, e.g.][]{grat01,rc02,rc03}. Low-resolution spectroscopic analyses (both for RGB and MS-SGB stars) were able to trace large spreads and anti-correlations in the strength of CN and CH bands, in some cases with clearly bimodal distributions
\citep[see, e.g.][for recent results and references]{sara,k08,pan10,lardo12a,lardo12b}.

On the other hand, the splitting of the MS detected in the Color Magnitude Diagram (CMD) of some clusters can be explained only with the
presence of a large intra-cluster spread in He abundance \citep[e.g.][]{norris,DAn05,piotto07}. 
The RGB bump has also been used as a tracer of He spread within clusters \citep{angie,nataf}.
The splitting of
SGB has also been observed in several clusters, although a univocal interpretation for this specific phenomenon is still lacking \citep[see][]{mi08,milo_tuc,piotto12}. 

Multi-band photometry involving a near UV passband encompassing the wavelength range $3000\AA \la \lambda\la 4000\AA$ 
(e.g. Landolt's U band, Str\"omgren $u$ band, F336W filter on the Hubble Space Telescope cameras) 
allowed the discovery that RGB of GCs appears split and/or 
broader than that expected from observational errors, with clear correlation between RGB colors at a given magnitude and abundance of Na or N
\citep[][]{yong,euge6752tut,euge3201,krav,fabi_m4,lardo,lardo12b}. It has been convincingly demonstrated that colors including near UV filters trace the strong variations of the strength of NH, CN, CH spectral features lying in that region of spectra of RGB stars \citep{euge11,sbor,milo_tuc}.
In particular, \citet[][Mi12 hereafter]{milo_tuc} have shown very clearly that the combination of spreads in the abundances of He, C, N, and O are responsible of a variety of photometric features in the CMD of GCs hosting multiple populations, depending on the considered evolutionary sequence and the set of adopted passbands. 

Interestingly, stars belonging to different generations (i.e. to different episodes
of star formation) appear well separated in color along a given evolutionary sequence on the CMD 
(e.g., along the MS or the RGB; see Fig.~35 by Mi12) and show systematic differences in their radial distributions. 
It has been generally found  that second generation stars (rich in N, He, and Na) are more centrally concentrated than the 
first generation stars \citep[see][and G12 for review and references]{lardo,vespe}.
At present, IC4499 is the only Galactic GC with accurate near UV photometry that does not show the
photometric  signatures of multiple populations \citep[][]{4499}.

In \citet[][hereafter L11]{lardo} we used public u,g,r photometry of Galactic GCs obtained by 
\citet{an} from Sloan Digital Sky Survey \citep[SDSS, see][]{dr7} images, to look for near-UV color spread along the RGB. 
Seven of the nine surveyed clusters showed statistically significant color spread while for the remaining two (NGC~5466 and NGC~2419)
the available photometry was not sufficiently accurate to detect any sign of spread.
The correlation between u-g color at a given magnitude along the RGB and Na abundance of individual stars was also demonstrated, at least in one case, and the significant differences in the radial distribution of blue and red RGB stars were found in all seven GCs with positive detection of the color spread. A deeper photometric investigation of NGC~2419 was a natural follow up of the analysis by L11.

%In the present paper we describe the analysis of these data and the (positive) results we have obtained. It is important to remark that, given the limitations inherent to the available dataset, in L11 we focused on the detection of significant differences between the color spread about a RGB ridge-line seen in purely optical colors (g-r) and that seen in optical-near-UV colors (u-g): the fact that the latter was significantly larger than the former, even when normalized to the respective photometric errors, was taken as the signature of the spectral anomalies associated with multiple populations. In the present case we had the opportunity (a) to design observations optimized for our scientific goal, i.e. many high signal-to-noise (S/N) exposures per filter under good seeing conditions, (b) to have under full control the whole data reduction procedure, that was also optimized for our main goal, and, as a consequence, (c) to perform extensive artificial stars experiments to have a full and accurate characterization of all the observational effects, including blendings. 

NGC~2419 is a metal-poor ([Fe/H]=-2.1, \citealt{judy12,muccia}), very luminous \citep[$M_V=-9.5$,][]{mic12} and massive \citep[$M\simeq 10^6 M_{\sun}$,][]{iba11,iba12} GC located in the far outskirts of the Milky Way \citep[$R_{GC}=$94.7~kpc, adopting the distance estimate by][]{dic_dist}.
Its half-light radius \citep[$r_{h}\sim 24$ pc][]{mic,iba11}\footnote{This is the 3-D half-light  radius, while the projected (2-D) half-light radius is $\sim 15$~pc.} is by far larger than that of other GCs of the same luminosity and is more akin to the nuclei of dwarf galaxies than to classical GCs \citep[][]{mack,brodie,new03,cas09}\footnote{See \citet{sippel} for an alternative view on the origin of large $r_h$ in distant GCs.}.
Another rare feature that NGC~2419 shares with other peculiar clusters (e.g., $\omega$~Cen, M54) is the presence of a conspicuous Blue Hook population, 
at the hottest extreme of the Horizontal Branch~\citep[HB;][]{ema}.

In spite of all the above hints suggesting the intrinsic complexity of this object,
high quality spectroscopic and photometric observations able to unveil the subtle effects of multiple populations in this cluster were lacking
until very recent times. \citet[][hereafter D11]{dic_mult} assembled accurate photometry from different telescopes, including Hubble Space Telescope (HST) and in different passbands, yet lacking near-UV ones. In spite of this, they were able to demonstrate that the observed F435W-F814W  color spread on the RGB was significantly larger than that expected from observational errors\footnote{The same result was found also for F457W-F850LP and F55W-F775W colors.}. They found that the observed RGB color distribution can be consistently interpreted together with the peculiar HB morphology of the cluster by assuming that $\sim$30\%  of the cluster stars belong to a second generation that is heavily enriched in He (D11 provide Y=0.42 as an indicative, reference value). D11 show that at the very low metallicity of NGC~2419 such a high He abundances is the prevailing factor that determines the color of RGB stars of the second generation that, at odds with other cases described above, lies {\em to the blue} of first generation stars, at least in the colors they considered and near the base of the RGB, where the color spreads are more evident (see L11, discussion and references therein).

\citet[][hereafter Mu12]{muccia} obtained abundances of Fe, Mg, K, Ti and Ca for 49 RGB stars of NGC~2419 from high 
S/N medium resolution spectra. They found that (a) the observed iron (as well as Ti and Ca) abundances are consistent with {\em no intrinsic spread}, (b) both Mg and K abundances display huge spreads, $-1.2\la$[Mg/Fe]$\la+1.0$ and $-0.2\la$[K/Fe]$\la+2.0$, and obvious bimodal distributions, and, finally, (c) [Mg/Fe] and [K/Fe] are strongly anti-correlated. Such large Mg and K spread as well as the anti-correlation between these two elements were never observed before, neither in GCs or in dwarf galaxies. These results have been nicely confirmed by \citet{judy12} from the analysis of high-resolution spectra of 13 bright RGB members. The latter also found a significant spread in Na abundance among the 
sampled stars, in particular among the eight K-poor (and Mg-rich) ones.

Mu12 noted that the large spread in Mg at constant Fe abundance and the similarity in the fraction of Mg-deficient stars and of putative He-rich stars \citep[according to][]{dic_mult} suggest that the observed abundance pattern may trace an extreme case of the same self-enrichment process due to multiple populations that is at work in other GCs. Indeed, a very recent paper by \citet{ventura}, proposes a theoretical framework based on self-enrichment from massive ($M\sim 6 M_{\sun}$) Asymptotic Giant Branch (AGB) stars that appear as a significant step forward in explaining the many peculiar features observed in this cluster. 

%The structural properties of NGC~2419 makes it one of the rare GCs where the two-body relaxation is pretty ineffective in settling energy equipartition \citep{baum09}. Indeed, \citet{ema} and \citet{mic12} independently showed that there is no sign of mass segregation among cluster stars. This is especially relevant in the present context, since any difference in the radial distribution between stars of different generations {\em must reflect}, in this case, the initial conditions set up at the cluster birth, without any significant contribution from secular dynamical evolution driven by two-body encounters.

In this paper we use high quality near-UV and optical photometry obtained with the Large Binocular Camera (LBC) mounted at the Large 
Binocular Telescope (LBT) to search for any intrinsic UV spread along the RGB stars of NGC~2419. This would provide additional and direct
observational support to the notion that the self-enrichment process typical of GCs also occurred in this cluster.

\section{Observations and data reduction}
\label{phot}

\begin{figure}
\includegraphics[width=80mm]{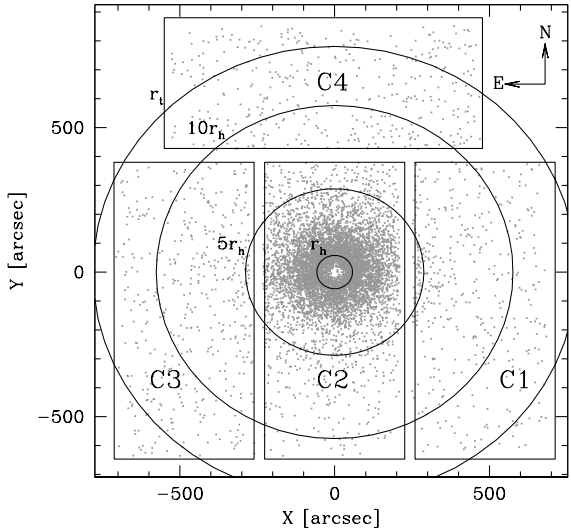}
 \caption{Map of the stars included in our final photometric catalog. The FoV of the various chips of the LBC camera (rectangular regions) are labelled as C1, C2, C3, C4. Circles centered on the cluster center and with radius 1$r_h$, 5$r_h$,  10$r_h$, and 1$r_t$ are superposed, for reference \citep[$r_h$ is the half-light radius, $r_t$ is the tidal radius, from the best-fit model (\# 17) by][]{iba11}.
 }
 \label{mapsel}
\end{figure}

Photometry was acquired at the LBT (Mount Graham; AZ), on the night of October 23, 2011, under good seeing conditions ($0\farcs7-0\farcs9$)
using the blue and red channels of the LBC \citep[LBC-B and LBC-R, respectively;][]{lbc}, simultaneously.
The optics of each LBC camera feeds a mosaic of four 4608~px~$\times$~2048~px CCDs, with a pixel scale of 0.225 arcsec~px$^{-1}$. Each CCD chip covers a field of $17.3\arcmin\times7.7\arcmin$. Chips 1, 2, and 3  are flanking one another, being adjacent along their long sides; Chip 4 is placed perpendicular to this array, with its long side adjacent to the short sides of the other chips \citep[see Fig.~4 of][]{lbc}. During our observations the pointing was chosen to place the center of NGC~2419 at the center of Chip~2 (see Fig.~\ref{mapsel}). 
In the following, we will use the terms Chip~1,2,3,4 referring both to the chips themselves and to the Field of View (FoV) they sample in the present case.

We used LBC-B to acquire 14 $t_{exp} = 300$~s exposures in the so called $u_{spec}$ filter, that reproduces the Sloan Digital Sky Survey (SDSS) u passband \citep[see][and references therein]{stripe82}. In the same time we used LBC-R to acquire 14 $t_{exp} = 90$~s exposures in V and I bands.

A photometric catalog of stellar magnitudes was obtained using an accurate Point Spread Function (PSF) fitting
procedure performed through DAOPHOTII~\citep[][]{s87}. Up to 60 well sampled and isolated stars in each
individual frame were chosen to model the PSF. We used a Moffat analytic function and a third-order look-up table was necessary in order 
to properly account for the spatial variation of the PSF~\citep[see][and http://lbc.mporzio.astro.it/commissioning/psf.html]{lbc}.
A master list of stars was obtained using all the stars detected in at least 6 of the 14 V-band images.
This approach delivers a master catalog free of spurious detections, such as cosmic rays or haloes and spikes around saturated stars.

The resulting master list was then used to perform PSF fitting over the entire dataset, adopting the standard ALLFRAME routine~\citep[][]{s94}.
Finally, the average of the magnitudes of the stars measured in at least 10 of the 14 frames of each 9band was adopted 
as the stellar magnitude in the final catalog, and the error on the mean was assumed as the associated photometric uncertainty.
The final catalog contains a total of $\sim11,500$ stars sampled in the $u$, V and I bands.

%%%%%%%%%%%%%%%%%%%%%%%%%%%%%%%%%%%%%%%%%%%%%%%%%% FIG 
%%%%%%%%%%%%%%%%%%%%%%%%%%%%%%%%%%%%%%%%%%%%%%%%%%%%%%%%%%%%%%%%%%%%%%%

\subsection{Photometric calibration, astrometry, and sample selection}
\label{cal}

%%%%%%%%%%%%%%%%%%%%%%%%%%%%%%%%%%%%%%%%%%%%%%%%%% FIG 
\begin{figure}
\includegraphics[width=80mm]{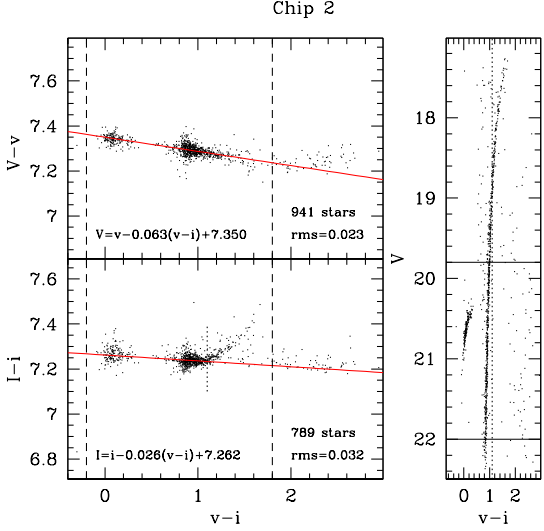}
 \caption{Left panels: standard system V(I) magnitudes minus instrumental v(i) magnitudes as a function of v-i color for the stars in common between our catalog (chip 2 only) and the catalog of secondary standard stars by 
\citet{s00,s05}. The vertical dashed lines enclosed the range of instrumental colors covered by cluster stars. The red solid lines display the adopted calibrating equations. The dotted segment marks the red limit beyond which a single linear relation is no more adequate to calibrate
all the stars. The same limit is marked as a dotted line in the right panel, showing the CMD of the stars in common between the two catalogs. The solid horizontal lines enclose the magnitude range of the RGB stars that are the object of our analysis.}
 \label{calib}
\end{figure}
%%%%%%%%%%%%%%%%%%%%%%%%%%%%%%%%%%%%%%%%%%%%%%%%%%%%%%%%%%%%%%%%%%%%%%%

%%%%%%%%%%%%%%%%%%%%%%%%%%%%%%%%%%%%%%%%%%%%%%%%%% FIG 
\begin{figure}
\includegraphics[width=80mm]{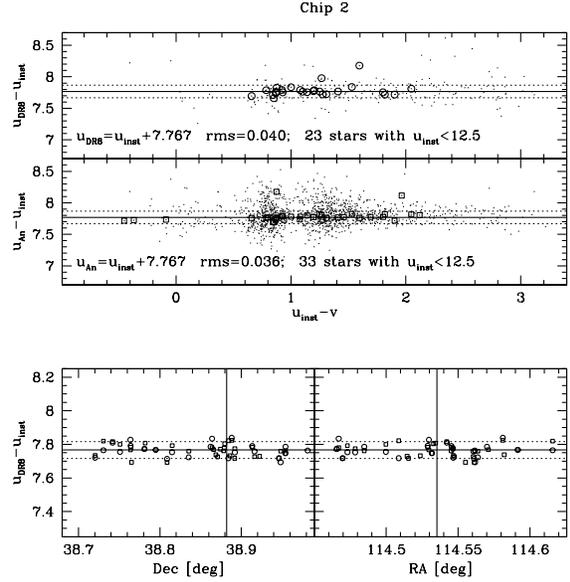}
 \caption{Upper panels: SDSS calibrated u magnitudes minus instrumental magnitudes $u_{inst}$ as a function of instrumental $u_{inst}$-v color for stars from SDSS-DR8 (upper panel) and from An et al. (middle panel) in common with our photometric catalog (chip2 only). The solid lines are the color-independent ZP we adopted to calibrate our instrumental u photometry; the dotted lines enclose a range of $\pm 0.1$ mag around the solid line. Open circles (upper panel) and open squares (middle panel) highlight the brightest stars ($u_{inst}<12.5$, corresponding to $u=20.27$, above the RGB tip) that have been used to derive the Zero Point; those lying beyond the $\pm 0.1$ mag range around the mean have been discarded. Lower panels: the standard-vs-instrumental magnitude difference for the same stars (with the same symbols) is plotted versus Dec and RA. The dotted lines enclose a range of $\pm 0.05$ mag around the solid line, marking the ZP level; the solid vertical lines marks the position of the cluster center.}
 \label{calibu}
\end{figure}
%%%%%%%%%%%%%%%%%%%%%%%%%%%%%%%%%%%%%%%%%%%%%%%%%%%%%%%%%%%%%%%%%%%%%%%

The transformation from instrumental to calibrated magnitudes was performed using large numbers of secondary standard stars that were included in our FoV and were successfully measured and included in our final catalog. For all the considered passbands (a) we derived a transformation for Chip 2, that contains most of the standard stars and will be the main focus of our analysis, (b) we applied the same transformations to stars from all the LBC chips, then (c) we used the secondary standards in each chip to adjust the photometric Zero Points (ZP). In this way the general calibrating equations are derived from the largest sample of standard stars and any chip-to-chip photometric shift is corrected for using local standards.

V and I magnitudes were calibrated using the large set of accurate secondary standards by \citet{s00,s05}. In the left panels of Fig.~\ref{calib} the solutions for Chip~2, as derived from more than 700 standard stars per filter, are presented. The dashed lines enclose the range in instrumental color covered by cluster stars. The derived first-order polynomials are roughly adequate in this range, and fully adequate for v-i$\le 1.1$. Since the stars being the main focus of the following analysis lie in this range (i.e. the RGB stars enclosed between the two solid lines in  the right panel of Fig.~\ref{calib}), the adopted transformations are appropriate for our purpose. It is interesting to note that a non-linear trend of I-i with color appears for cluster stars redder than v-i$=1.1$. This suggests that the differences between the actual LBC-R I filter and the standard I filter makes up a kind of ``color'' that is likely sensitive to stellar surface gravity, since bright cluster giants and foreground M dwarfs seem to have different behaviors as a function of v-i color.

We used the large number of standard stars in the FoV to verify the lack of any  significant residual trend (at the level of $\la 0.01$ mag) of the photometric ZP with position within Chip~2.

The ZP of the other chips were adjusted adding the following shifts ($\Delta ZP$):\\
Chip 1: $\Delta ZP(V)=0.00$ and $\Delta ZP(I)=+0.02$ from 58 standard stars in the FoV;\\
Chip 3: $\Delta ZP(V)=+0.02$ and $\Delta ZP(I)=+0.05$ from 63 standard stars in the FoV;\\
Chip 4: $\Delta ZP(V)=-0.01$ and $\Delta ZP(I)=+0.07$ from 21 standard stars in the FoV.

To calibrate u photometry we used as secondary standards the stars in common between our photometry
and the SDSS Data~Release~8 \citep[DR8,][]{dr8} public catalog of stellar sources. DR8 is the first SDSS release whose calibrated photometry fully relies on the so-called \"uber-calibration procedure \citep{ubercal}, achieving 1 per cent relative calibration errors in g,r,i,z and 2 per cent in the u band over the entire survey area. Since in the actual DR8 catalog both the number of stars with valid photometry and the photometric accuracy are limited by data reduction approach adopted by the SDSS in the most crowded area around the cluster center, we adopt the catalog by \citet{an} as an additional (and not independent) dataset to check and extend our calibration. \citet{an} re-reduced SDSS images using state-of-the art techniques for stellar photometry in crowded regions, obtaining catalogs with larger number of stars and with better relative photometry accuracy near the cluster center with respect to DR8. Since \citet{an} photometry was based on DR7 calibrations we adjusted their ZP to the new \"uber-calibrated DR8 system using 1655 stars in common between the two catalogs\footnote{The adopted transformation is $u_{DR8}=u_{An}-0.018$.}.

In the upper panels of Fig.~\ref{calibu} we show the adopted solution for the calibration of u photometry compared with the two datasets; only stars brighter than u$\simeq 20.3$ were adopted to fit the ZP (open symbols) that is clearly independent of color over the wide covered range. The comparison with the more extended (in number and color) catalog by \citet{an} fully confirms the adequacy of the solution. In the lower panels it is shown that the adopted best (bright) standards do not trace any obvious trend of the photometric ZP(u) with position in the chip. However the considered standard stars are not so many and the scatter is larger than in V and I: trends with position of maximum amplitude $\la 0.03$ mag across Chip~2 FoV can, in principle, be present and go unnoticed. {\em This is a caveat that should be kept in mind in the following}. While it is quite unlikely that such an undetected trend is at the origin of the results of our analysis, we do not have any real control on its actual effects. 

More than two hundred local standards per chip were used to adjust the ZP of chip 1, 2, and 4.
The derived shifts are null, except for Chip 4, where $\Delta ZP(u)=+0.03$.

To minimize any subtle residual photometric trend as a function of position that may affect our results, {\em we will limit the core of our analysis only to stars lying in Chip~2} (see Sect.~\ref{csprea}). Fig.~\ref{mapsel} shows that this allow us to maintain a good radial coverage (from the cluster center out to $\ga 10r_h$) while dealing with the best calibrated and cross-checked photometric sample. Photometry from other Chips have been used only to estimate the impact of foreground contamination of our CMDs (see Sec.~\ref{csprea}). We provide a detailed description of the whole procedure of calibration as a reference for future users of our photometric catalog that we make electronically available~(a small sample is shown in Table~\ref{tab_cat}) .

\subsubsection{Astrometry}

The star coordinates were transformed from X,Y in pixel to Equatorial J2000 RA and Dec, using SDSS stars as astrometric standards, with the code CataXcorr\footnote{Developed by P. Montegriffo, together with a suite of other astronomical utility softwares (CataPack), at INAF-OABo and extensively used by several authors since more than ten years.}. The transformation was performed in two steps:

\begin{itemize}

\item first, through a third order polynomial using the DR8 catalog, that covers uniformly the whole LBC FoV except for the immediate surroundings of the cluster center (because of crowding).

\item then, through a first order polynomial using the \citet{an} catalog that is much more populated near the cluster center but misses the corners of our FoV.

\end{itemize}

The final solutions has a r.m.s of 0.04$\arcsec$, both in RA and Dec, from $\sim 2400$ stars in common (Chip 2). The difference in position between our astrometric system and that by \citet{s05} has a r.m.s of $0.04\arcsec$ in RA and $0.05\arcsec$ in Dec for 2745 stars in common. The r.m.s. is always $\le 0.07\arcsec$ also for all the other chips, both with respect to DR8 or \citet{s05} astrometry, albeit with significantly less astrometric standards (from $\sim 100$ to $\sim 400$, depending on the case).

In the following we will adopt the coordinates of the cluster center by \citet{ema} and the estimate of the half-light radius $r_h=56.3\arcsec$ by \citet{iba11}, as a reference lenght-scale. 

\subsubsection{Sample selection}
\label{sel}

The scientific goal of our experiment requires stringent quality control: for our purpose it is much better to lose stars whose photometry is somehow uncertain than including spurious sources or measures. For this reason we adopted pretty conservative criteria on the quality parameters provided by DAOPHOT, i.e. $CHI$ and $SHARP$ \citep[see][]{s87,s94}. By simultaneous inspection of the CMD and the $SHARP$ vs. V and $CHI$ vs. V diagrams we finally adopted the following magnitude-dependent criteria. We accept as bona-fide well measured stars for the following analysis sources having

\begin{enumerate}

\item $|SHARP|<0.1$ if V$<20.0$

\item $|SHARP|<0.2$ if $20.0\le$V$<23.0$

\item $|SHARP|<0.4$ if V$\ge 23.0$

\item and $CHI<2.0$ for any V

\end{enumerate} 

%Since the distribution of the original {\em sharpness} values ($SHARP_o$, as provided by DAOPHOT) was not peaked at 0.0 we adopted the symmetric selection criteria above on $SHARP=SHARP_o-0.03$. 

The above criteria select a sample of 8036 sources, 6802 of them in Chip~2. The following analysis will always deal with this selected sample, even if we maintain all the original 11495 sources in our catalog.
In Sect.~\ref{arti}, below we will introduce a further selection based on the distance from the cluster center, i.e. we will exclude from our analysis stars with $R\le 50\arcsec$. A sample of the photometric catalog of selected stars in the entire mosaic is presented in Table~\ref{tab_cat}.

\begin{table*}
\label{riass}
 \centering
  \caption{A sample of the photometric catalog of NGC~2419. The complete catalog is
available in ASCII format in the electronic edition of the paper.}
  \begin{tabular}{@{}lcccccccc@{}}
  \hline
ID & $\alpha_{2000}$&$\delta_{2000}$ &
u &$\epsilon_u$ & V & 
$\epsilon_V$ & I & $\epsilon_I$\\
 \hline
 100001 & 7$^h$ 37$^m$ 47.06$^s$ & 38$\degr$ 46$\arcmin$ 03.34$\arcsec$  & 24.322 & 0.037 & 20.873 & 0.004 & 19.089 & 0.003\\
 100003 & 7$^h$ 37$^m$ 46.97$^s$ & 38$\degr$ 50$\arcmin$ 23.79$\arcsec$  & 24.330 & 0.040 & 21.367 & 0.005 & 19.426 & 0.003\\
 100004 & 7$^h$ 37$^m$ 46.66$^s$ & 38$\degr$ 45$\arcmin$ 40.01$\arcsec$  & 23.689 & 0.017 & 22.978 & 0.019 & 21.955 & 0.015\\
 100005 & 7$^h$ 37$^m$ 46.78$^s$ & 38$\degr$ 50$\arcmin$ 02.51$\arcsec$  & 21.320 & 0.004 & 19.259 & 0.003 & 18.168 & 0.003\\
 100008 & 7$^h$ 37$^m$ 46.66$^s$ & 38$\degr$ 50$\arcmin$ 19.75$\arcsec$  & 21.448 & 0.005 & 19.392 & 0.002 & 18.312 & 0.002\\
 100009 & 7$^h$ 37$^m$ 46.79$^s$ & 38$\degr$ 53$\arcmin$ 47.15$\arcsec$  & 23.292 & 0.012 & 21.764 & 0.007 & 20.841 & 0.005\\
 100011 & 7$^h$ 37$^m$ 46.62$^s$ & 38$\degr$ 52$\arcmin$ 01.63$\arcsec$  & 23.759 & 0.021 & 23.060 & 0.028 & 22.470 & 0.032\\
 100012 & 7$^h$ 37$^m$ 46.56$^s$ & 38$\degr$ 51$\arcmin$ 25.49$\arcsec$  & 24.498 & 0.035 & 23.284 & 0.026 & 22.566 & 0.022\\
 100014 & 7$^h$ 37$^m$ 46.84$^s$ & 38$\degr$ 55$\arcmin$ 42.20$\arcsec$  & 22.212 & 0.005 & 19.040 & 0.002 & 17.491 & 0.003\\
 \hline
\end{tabular} \label{tab_cat}
\end{table*}

\subsection{The Color Magnitude Diagrams}
\label{cmd}

%%%%%%%%%%%%%%%%%%%%%%%%%%%%%%%%%%%%%%%%%%%%%%%%%% FIG 
\begin{figure*}
\includegraphics[width=120mm]{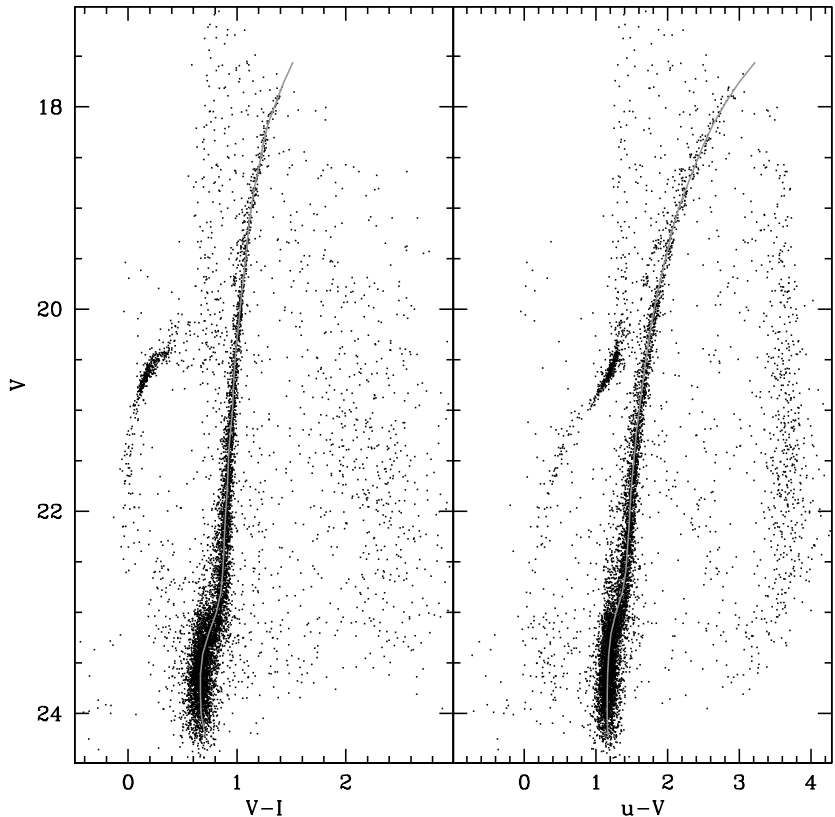}
 \caption{V,V-I (left panel)and V,u-V CMD for all the stars included in our final photometric catalog (from all the four LBC chips). The 
 MS and RGB  ridge lines are also shown (grey line).
 }
 \label{cmdall}
\end{figure*}
%%%%%%%%%%%%%%%%%%%%%%%%%%%%%%%%%%%%%%%%%%%%%%%%%%%%%%%%%%%%%%%%%%%%%%%

In Fig.~\ref{cmdall} we present the V, V-I and V, u-V Color Magnitude Diagrams of our selected sample (all chips). The well defined cluster RGB is the dominant feature, from $V\sim 18$ down to $V\sim 23$ where it bends into the SGB and MS down to a limiting magnitude $V\simeq 24.5$. The prominent blue HB sequence is visible from $V\simeq 20.3$ to $V\sim 22.5$. We verified, by cross-matching with the deeper Subaru and HST photometry by \citet{ema}, that the handful of faint blue stars at $V-I\simeq u-V\la 0.0$ and $V<23.0$ are extreme BHB stars \citep[blue-hook,][]{ema,dic_mult}. The majority of the sources at $V<20.5$ and u-V$<1.0$ are not among the variable stars listed by \citet{dic_dist}, but have, instead the colors typical of quasars. Indeed a few of them have been counter-identified as known quasars. We are following up these sources since they can provide an excellent reference frame for the determination of the absolute proper motion of the cluster.

The vertical stripe of stars with a sharp color-edge at $V-I\sim 0.8$ and $u-V\simeq 1.2$ is made of foreground Turn-Off stars at various distances in the Galactic Halo. For  $V-I\ga 2.0$ and around $u-V\sim 3.6$ the plume of Galactic M-dwarfs is clearly visible. In the V, u-V CMD a hint of the MS of the Monoceros Stream \citep[lying in the foreground, see][]{sol}, can be also discerned, bending from (V, u-V)$\sim$(21.0,1.0) to (V, u-V)$\sim$(23.5,3.0).

Several stars within $\sim 0.5$ magnitudes from the tip of the cluster RGB (at $V\simeq 17.4$) have been excluded from the selected sample by the criteria described in Sect.~\ref{sel}, likely because they were partially saturated. 
%For this reason the typical photometric uncertainty for such stars is larger than expected at such % magnitude. 
Finally, in Fig.~\ref{cmdall} we superpose to the cluster MS and RGB the ridge lines we derived from the data by averaging the color (with 2-$\sigma$ clipping) over magnitude bins of variable width. These ridge lines are always used as a reference in the following.

We do not see any obvious sign of differential reddening across our FoV. This is in good agreement with D11 who set a strong upper limit of 0.005 mag to differential reddening over the central field of view sampled by their HST data.

\subsection{Artificial stars experiments}
\label{arti}

As previously discussed, the aim of this work is to search for color spreads along the RGB. 
It is then crucial to have a realistic and robust estimate of any possible unphysical factors 
that can induce the broadening of the RGB~\citep[e.g. blending, photometric errors; see][]{lardo}.

These features, related to the quality of the data, can be properly studied through artificial stars experiments.
We produced a catalog of artificial stars following the procedure described in~\citet[][]{mic02}.
We first generated a catalog of simulated stars with a V magnitude randomly extracted from a Luminosity Function (LF) 
reproducing the observed LF in the V band. The $u$ and I magnitudes were assigned at each sampled V magnitude 
by interpolating the mean ridge lines of the cluster (see Fig.~\ref{cmdall}).

The artificial stars where then added to the real images
using DAOPHOTII/ADDSTAR routine and adopting the same PSF models computed during the PSF fitting of the images (see Sect.~\ref{phot}).
In order to avoid the risk to induce artificial stellar crowding, each frame was divided in a grid of boxes of the size 
of 20 pixel (i.e. 5 times the mean FWHM of the stars in the frames) and only one star was randomly 
placed within each box in each artificial test run. 
Once the artificial stars were added on the images, we performed the photometric reduction adopting exactly the same
approach described in Sect.~\ref{phot}. The entire procedure
was repeated many times and we collected a final catalog of more than 200,000 artificial stars. 
We finally applied to the artificial stars the same selection criteria used for real stars (see Sect.~\ref{sel}).

%%%%%%%%%%%%%%%%%%%%%%%%%%%%%%%%%%%%%%%%%%%%%%%%%% FIG 
\begin{figure}
\includegraphics[width=80mm]{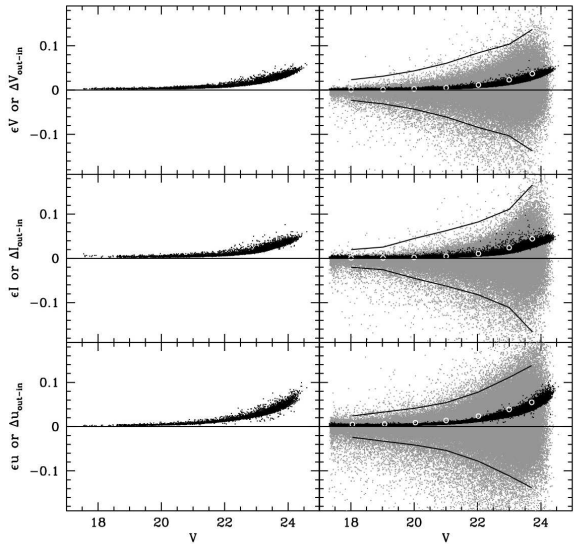}
 \caption{Black dots: photometric errors ($\epsilon_{\lambda}$, where $\lambda$= V, I, u) from repeated measures of real (left panel) and artificial (right panel) stars as a function of V magnitude. Grey dots: differences between input and output magnitudes for artificial stars; the solid lines are $\pm 3\sigma$ limits around $\Delta\lambda_{out-in}=0.0$. Open grey circles are the r.m.s. of $\Delta\lambda_{out-in}$ computed in bins of one magnitude from stars enclosed within the $\pm 3\sigma$ lines.
 }
 \label{errors}
\end{figure}
%%%%%%%%%%%%%%%%%%%%%%%%%%%%%%%%%%%%%%%%%%%%%%%%%%%%%%%%%%%%%%%%%%%%%%%

In Fig.~\ref{errors} we show the photometric errors computed as errors on the mean from repeated measures (black dots), both for real (left panels) and artificial stars (right panel). It is reassuring that the error distributions in the two sets are fully consistent, and that the r.m.s. of the differences between input and output magnitudes has also a similar behaviour (grey circles). %This support the validity of the approach introduced by L11, i.e. the normalisation of the color %spread to the photometric error. 
Note that the average photometric error is $\le 0.01$ mag in all the considered passbands for RGB stars in the magnitude range that we consider for our analysis ($19.8\le V\le 22.0$, see Sect.~\ref{csprea}, below).

\begin{figure}
\includegraphics[width=80mm]{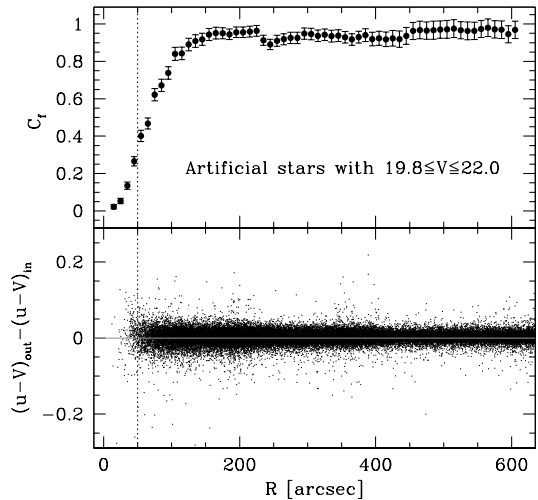}
 \caption{Upper panel: completeness as a function of distance from the center of the cluster for stars in the same range of magnitude as the RGB stars that are the main object of our analysis.
  Lower panel: distribution of the differences between input and output u-V colors for stars in the same magnitude range. The dotted lines mark the $R=50\arcsec$ threshold.}
 \label{comple}
\end{figure}

In Fig.~\ref{comple} the completeness and the scatter in u-V color are plotted as a function of distance from the cluster center for the same sub-set of (artificial) RGB stars. The completeness within $R\le 50\arcsec$ is so low that in the following we exclude stars in this radial range from our analysis. For $R\ga 150\arcsec$ the completeness is constant with radius. The small discontinuity at $R\sim 230\arcsec$ is due to an additional loss of stars in a radius of $\sim 50\arcsec$ around the bright foreground star HD60771 (V=7.23).

In the following analysis we will made use of sub-samples of artificial stars having the same magnitude, color and radial distribution as the real sample we are considering in our analysis.
These samples are obtained by associating to each real star a successfully recovered artificial 
star lying in the immediate proximity and having similar magnitude and color of the real star, as done and described in detail in \citet{mic12}. We define these sets of artificial stars as {\em Similar Samples}. Similar Samples provide the best approximation of the effects of the observation + data reduction process as the {\em actual sample of real stars under consideration}. This is a safe approach when dealing with cases in which, for instance, the completeness and/or the photometric error distributions may change significantly with radius \citep[see Fig.~\ref{comple}, and][]{mic12}.
Notice that whenever we use a Similar Sample in our analysis, we counter-check our results adopting three different (albeit not fully independent) random realisations of the sample.

%%%%%%%%%%%%%%%%%%%%%%%%%%%%%%%%%%%%%%%%%%%%%%%%%% FIG 
%%%%%%%%%%%%%%%%%%%%%%%%%%%%%%%%%%%%%%%%%%%%%%%%%%%%%%%%%%%%%%%%%%%%%%%

\section{Color spread along the RGB}
\label{csprea}

\begin{figure}
\includegraphics[width=80mm]{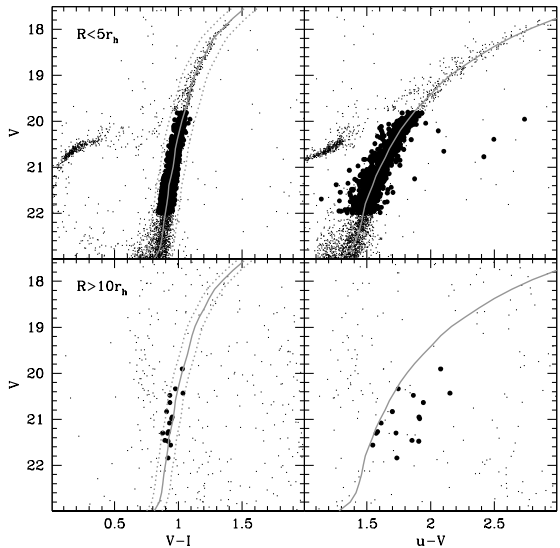}
 \caption{Upper left panel: the RGB stars to be searched for color spread (heavy dots) are selected in a narrow strip ($\pm 0.06$~mag) around the V, V-I ridge line (grey solid line; dotted lines enclose the $\pm 0.06$~mag color range). Upper right panel: the stars selected in V-I are plotted
 in the $u$-V, V CMD. To show the behavior of cluster stars we have plotted in the upper panels CMDs only stars lying within $5r_h$ from the cluster center. To provide an idea of the effect of field contamination on our analysis, in the lower panels we present the same CMDs for stars with $R>10r_h$ (from any Chip, see Fig.~\ref{mapsel}). 
 }
 \label{cmdsee}
\end{figure}

In the upper-left panel of Fig.~\ref{cmdsee} we show how we selected the RGB sample that is the object of the present analysis. We focus on stars on the lower RGB, below the RGB bump, where 
(a) the effects of CNO and He abundance differences on color spread are larger \citep[L11,Mi12][]{sbor}, (b) no effects of extra-mixing are expected \citep{grat00}, and (c) any possible
contamination from genuine AGB stars is avoided. We recall again that, in the following, we consider only stars from Chip~2 with $R>50\arcsec$, except for the lower panels of Fig.~\ref{cmdsee}. Following L11, we selected the most likely RGB candidates in V-I color, in a narrow ($\pm 0.06$ mag i.e. more than 4 times the combined photometric error in V an I bands) strip around the ridge-line.
Heavy points show how this selection translates into the u-V, V CMD in the upper-right panel of 
the figure. 

In order to have a robust estimate of the field contamination possibly affecting our RGB sample, we counted
the number of stars located at a distance $R>10r_h$ from the cluster center (from any Chip) and falling in the same selection box used
to our bona-fide RGB stars.
As shown in the lower panels of Fig.~\ref{cmdsee}, only 15 stars falls in this box. Since the ratio of the area of the two considered regions is $\sim 3.5$, less than 5 field stars are 
expected to contaminate the inner sample, where most of cluster RGB stars lie. This clearly demonstrates that fore/background 
contamination is negligible.

%%%%%%%%%%%%%%%%%%%%%%%%%%%%%%%%%%%%%%%%%%%%%%%%%% FIG 
%%%%%%%%%%%%%%%%%%%%%%%%%%%%%%%%%%%%%%%%%%%%%%%%%%%%%%%%%%%%%%%%%%%%%%%
%%%%%%%%%%%%%%%%%%%%%%%%%%%%%%%%%%%%%%%%%%%%%%%%%% FIG 
\begin{figure}
\includegraphics[width=80mm]{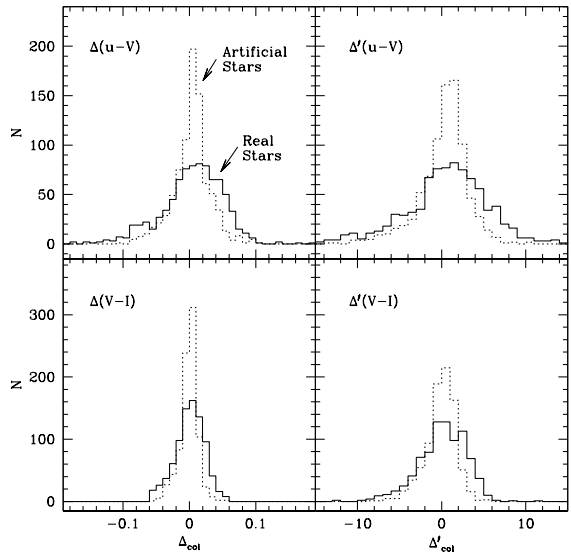}
 \caption{Distributions of u-V (solid histograms) and V-I (dotted histograms) color spreads with respect to the RGB fiducials of Fig.~\ref{cmdsee} for real stars of the selected RGB sample (lower panels) and for the corresponding sample of artificial stars with the same radial distribution (upper panels). Left panels show the distributions of the absolute color spread while right panels 
show the distributions of the {\em normalized} color spread, following L11.}
 \label{distri}
\end{figure}
%%%%%%%%%%%%%%%%%%%%%%%%%%%%%%%%%%%%%%%%%%%%%%%%%%%%%%%%%%%%%%%%%%%%%%%

Again following L11, we define the {\em absolute color spread} ($\Delta_{col}$; where col=V-I or u-V) as the difference (in magnitudes) between the color of a selected star and the ridge-line at the same magnitude, and
the {\em normalized color spread} as the same quantity divided by the corresponding photometric error ($\Delta_{col}\prime = \frac{\Delta_{col}}{\epsilon_{col}}$). 

In Fig.~\ref{distri} we compare the $\Delta_{col}$ and $\Delta_{col}\prime$ in V-I (lower panels) and u-V (upper panels) for
artificial and real stars (dotted and solid histograms, respectively). Artificial stars are selected in the same way as real stars (i.e. $19.8\le V\le22.0$ RGB stars within $\pm 0.06$ mag in color from the ridge-line in the V, V-I CMD) from a {\em Similar Sample}, having the same magnitude and radial distribution of real stars (and, by definition, the same number of stars, see Sect.~\ref{arti}). A few specific comment are in order:

\begin{enumerate}

\item In all cases the distributions of color spread in real stars are wider than in artificial stars, implying that an {\em intrinsic color spread is detected} both in V-I (in agreement with D11, who considered various combinations of optical colors) and in u-V (a new finding, not particularly surprising but not trivial, given the results by Mi12). Subtracting in quadrature the r.m.s. of the artificial stars distribution to that of real stars we find that the average intrinsic scatter is $0.018$~mag in V-I and $0.037$~mag in u-V. It turns out that the effect has its maximum amplitude in u-I, where the intrinsic spread is $0.082$~mag\footnote{We note that these numbers should be taken as measures of the {\em relative} sensitivity of the various colors to the same physical differences
between stars, since the distributions they are referred to are {\em not} Gaussian. The actual observed r.m.s. for real(artificial) stars are 0.0221(0.0133) mag,  0.0466(0.0284) mag,  and 0.0912(0.0395) mag,  in V-I, u-V, and u-I, respectively.}

\item In all the possible comparisons between color spread distributions for artificial and real stars shown in Fig.~\ref{distri} the probability that artificial and real stars are drawn from the same parent populations are always lower than $4\times10^{-15}$, according to Kolmogorov-Smirnov (KS) tests. The same is true for the spread in u-I (not shown in Fig.~\ref{distri}). {\em The detections of intrinsic color spreads in {\rm V-I}, {\rm u-V} and {\rm u-I} are highly significant}.

\item The fact that a significant color spread is detected not only in u-V and u-I \citep[i.e., in  optical-NUV colors, as in the majority of GCs, see][and L11]{yong} but also in V-I, provides support to the conclusions by D11, i.e. that in the case of NGC~2419 the dominant factor driving the color spread on the RGB, at least near the RGB base, is a large He spread, not the NH and CN bands, as the effects of these bands is weak in this very metal poor regime and it should be negligible in V-I (see L11, discussion and references therein). In Fig.~\ref{cmdhe} we use BASTI isochrones \citep{pi04} to show that the observed V-I color spread is roughly qualitatively consistent with the color difference induced by the difference in He abundance proposed by D11. 

\end{enumerate}

%%%%%%%%%%%%%%%%%%%%%%%%%%%%%%%%%%%%%%%%%%%%%%%%%% FIG 
%\begin{figure}
%\includegraphics[width=80mm]{ordine.ps}
% \caption{Upper panel: distribution of the absolute color spread in u-V, for the Blue (blue %solid histogram) and Red (red dashed histogram) RGB stars selected in u-V. Lower panel:
%distribution of the V-I color spread of the two RGB subsamples. According to a KS test, the %probability that the two samples are extracted from the same 
% parent distribution in V-I is$P=3.1\times10^{-18}$.
% }
% \label{ordine}
%\end{figure}
%%%%%%%%%%%%%%%%%%%%%%%%%%%%%%%%%%%%%%%%%%%%%%%%%%%%%%%%%%%%%%%%%%%%%%%%

\subsection{Radial distribution of Red and Blue RGB stars}

%%%%%%%%%%%%%%%%%%%%%%%%%%%%%%%%%%%%%%%%%%%%%%%%%% FIG 
\begin{figure}
\includegraphics[width=80mm]{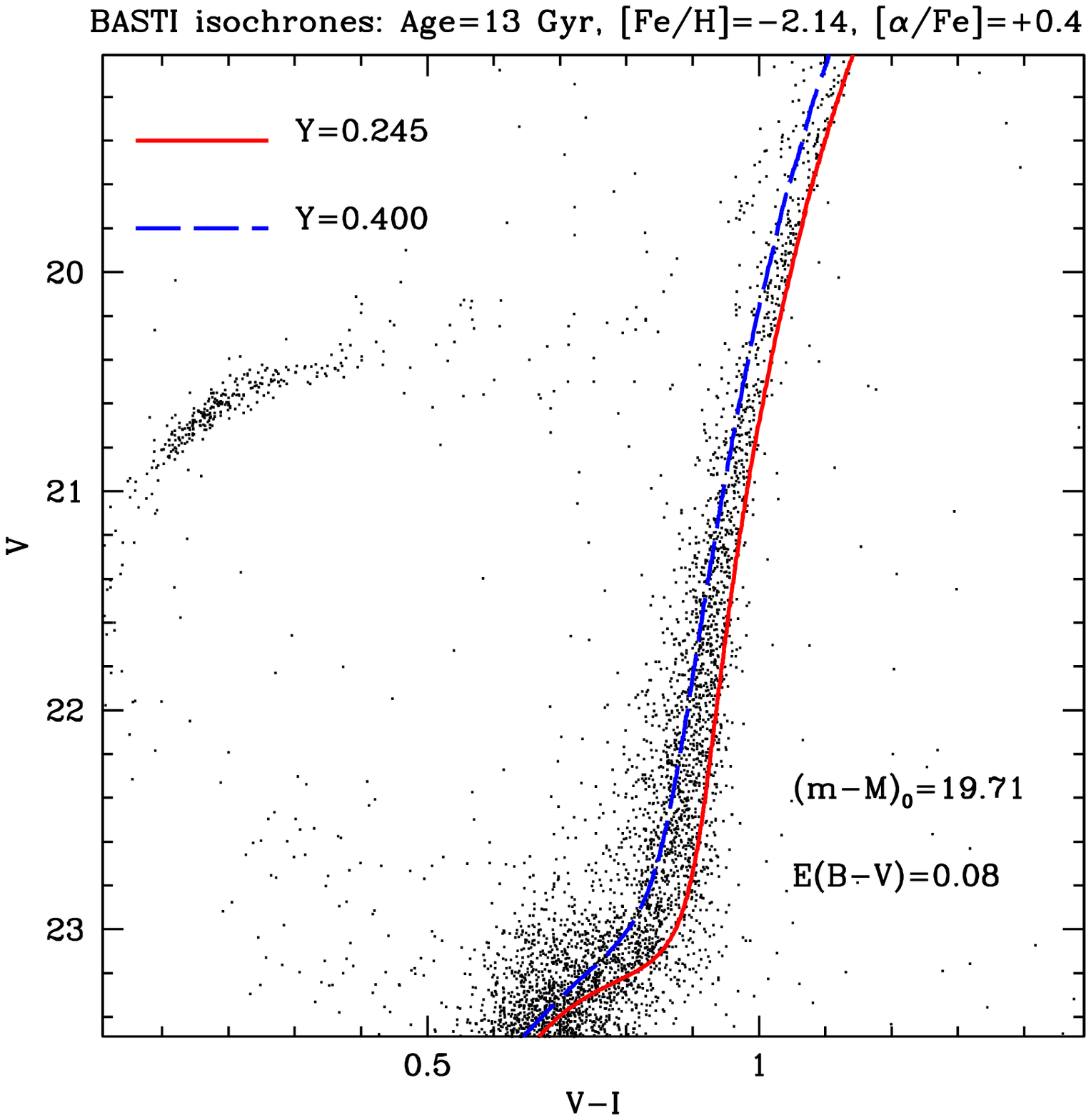}
 \caption{Theoretical isochrones from the BASTI dataset \citep{pi04} for two different He abundances are superimposed to the observed V, V-I CMD of NGC~2419. The age, metallicity and 
 [$\alpha$/Fe] ratio have been adopted following \citet{mic12}; the adopted distance modulus and reddening are from \citet{dic_dist}. The He abundance of the two model is similar to that assumed by 
D11 for the first (Y=0.245) and second generation (Y=0.400) of stars they propose.}
 \label{cmdhe}
\end{figure}
%%%%%%%%%%%%%%%%%%%%%%%%%%%%%%%%%%%%%%%%%%%%%%%%%%%%%%%%%%%%%%%%%%%%%%%

%%%%%%%%%%%%%%%%%%%%%%%%%%%%%%%%%%%%%%%%%%%%%%%%%% FIG 
\begin{figure}
\includegraphics[width=80mm]{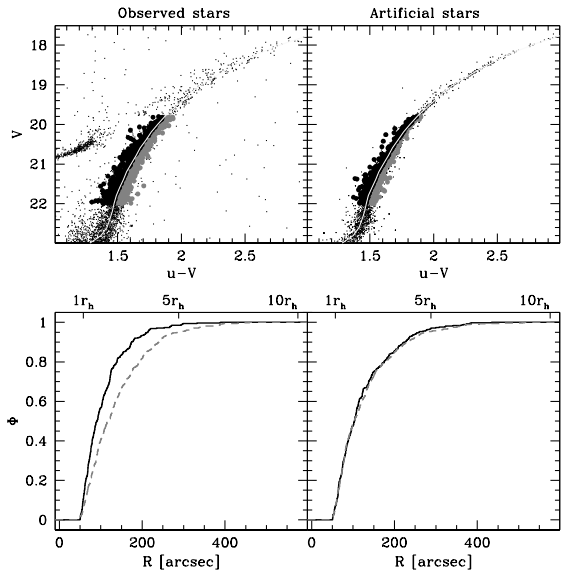}
 \caption{Upper panels: selection of Blue and Red RGB stars in the V, u-V CMD (left panel)
 and in the V, u-I CMD (right panel); the solid grey lines are the RGB ridge lines. Lower panels: 
 Comparisons between the radial distribution of Blue (black solid line) and Red (grey dashed line) RGB stars as selected in the corresponding CMDs. 
 }
 \label{radtut}
\end{figure}
%%%%%%%%%%%%%%%%%%%%%%%%%%%%%%%%%%%%%%%%%%%%%%%%%%%%%%%%%%%%%%%%%%%%%%%

We divide the RGB sample into RGB-Blue and RGB-Red according to the color of the stars with respect to the ridge-line. It is important to stress that, even with the high photometric quality of the data presented in this paper, we are
unable to detect any obvious split (bimodality) in color between the different populations along the RGB.
Any empirical separation between a RGB-Blue and a RGB-Red samples based on color is somewhat arbitrary. Even if the two populations were cleanly separated in terms of He and light elements abundance distributions, these physical differences would translate into color spreads whose amplitude changes with magnitude \citep[Mi12,][]{sbor}, and observational errors (that are also varying with magnitude) would contribute to mix the two populations. Hence any sample of stars lying to the blue(red) of the ridge-line in a given plane would contain a mix of the two parent populations (e.g., the first and second generation of stars in the cluster). Nevertheless,  we expect that each sample is {\em dominated} by the corresponding sub-population.

In the upper panels of Fig.~\ref{radtut} the selection of the RGB-Blue and RGB-Red sub-samples as perfomed in u-V and u-I (left and right panels, respectively) is displayed. In the lower panels the corresponding radial distributions of RGB-Blue and RGB-Red stars are compared. Independently of the adopted selection RGB-Blue stars are significantly more centrally concentrated than RGB-Red stars. According to a KS test the probability that the two sub-samples are extracted from the same parent radial distribution is $P_{KS}=2\times 10^{-9}$, for the (u-V)-selected samples, and $P_{KS}=8\times 10^{-11}$ for the (u-I)-selected sample\footnote{The same kind of difference is observed if the selection is performed in V-I, but in this case the significance is lower, $P_{KS}=0.058$. This is due to the lower discriminating power of the V-I color, the one displaying the smallest intrinsic spread, leading to a higher degree of mixing between {\em intrinsically} red and blue RGB stars in the resulting sub-samples.}. {\em This is the first time that a difference in the radial distribution between different samples of stars is detected in NGC~2419}.

In all the GCs where a difference in radial distribution has been observed between stars ascribable to different generations, second generation stars have been invariably found to be more centrally concentrated than first generation stars (see L11, G12 and Mi12 for references and discussion).
The result presented in Fig.~\ref{radtut} shows that NGC~2419 is no exception, if indeed RGB-Blue stars are dominated by second generation stars as envisaged by D11 and illustrated in Fig.~\ref{cmdhe}.

\begin{figure}
\includegraphics[width=80mm]{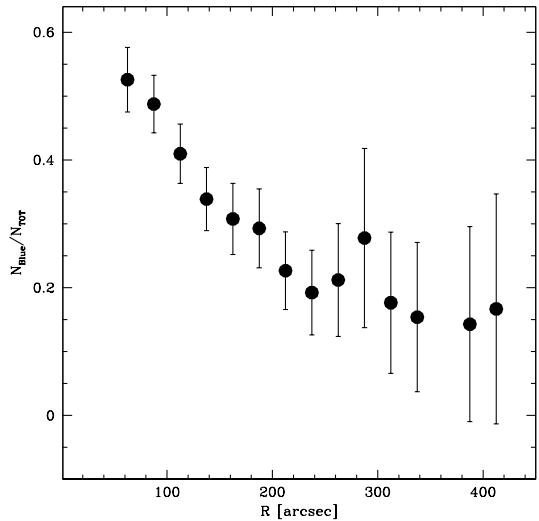}
 \caption{Fraction of RGB Blue stars (selected in u-I) on the total number of RGB stars in the selected magnitude range as a function of distance from the cluster center. The fraction is computed on running bins 50$\arcsec$ wide with a step of 25$\arcsec$. The profile obtained from 
 (u-V)-selected stars is indistinguishable.
 }
 \label{ratio}
\end{figure}

Fig.~\ref{ratio} shows the radial trend of the fraction of RGB-Blue stars, running from $\simeq$55 per cent at $R\sim 60\arcsec$ down to $\simeq$15 per cent at $R\sim 260\arcsec$, where the profile flattens, albeit with large (but not statistically significant) fluctuations. It is worth recalling that the $R\le 50\arcsec$ region is excluded from the analysis. The overall fraction of RGB-Blue stars within $10r_h$ is $\simeq 40$ per cent, in good agreement with the fraction of Mg-deficient stars derived by Mu12 and in rough agreement with the fraction of extreme He-rich stars as estimated by D11. 

In a recent paper \citet{vespe} demonstrated that the spatial segregation between first and second generation stars, set up at the birth of the second generation, can be preserved for a large fraction of the cluster life, depending on the original degree of segregation and on the rate at which 2-body relaxation is working in the cluster (but, in any case, for more than 10 half-mass relaxation times $t_{rh}$). As anticipated, in the case of NGC~2419 $t_{rh}$ is significantly larger than the Hubble time \citep{ema,harris}, hence the radial distributions of first and second generation stars should closely trace those settled at the end of the star forming phase \citep[see also][]{mic12}. It is interesting to note, from Fig.~\ref{ratio}, that the $N_{Blue}/N_{TOT}$ profile approaches the average global value ($\sim$ 40 per cent) at $R\sim 2r_h$, in good agreement with the predictions by \citet{vespe}.

An important test to verify the scenario by D11, where RGB-Blue stars should belong to the same He-enriched second generation as Extreme HB (EHB) stars, would be to search for a radial trend similar to that shown in Fig.~\ref{ratio} for EHB stars. Unfortunately, this  is quite difficult to probe because of the large magnitude difference between BHB and EHB, implying a significant difference in the completeness fraction between the two samples, and, above all, the difference in the radial trend of the completeness as a function of radius \citep[see Fig.~\ref{comple} and][]{mic12}. In particular \citet{mic12} showed that even with the deepest HST photometry, the completeness factor of stars as faint as EHB would be lower than 50 per cent for significant fractions of the radial range sampled  by the FoV of HST cameras, and in any case it is subject to strong variations with radius over the whole FoV. Hence, the considered test cannot be performed in a safe way with the observational material currently available (and, indeed, was not performed by D11). 
On the other hand, the results shown in Fig.~\ref{radtut} and Fig.~\ref{ratio} are fully reliable since the RGB-Blue and RGB-Red samples
are equally affected by incompleteness. 

%%%%%%%%%%%%%%%%%%%%%%%%%%%%%%%%%%%%%%%%%%%%%%%%%% FIG 
%%%%%%%%%%%%%%%%%%%%%%%%%%%%%%%%%%%%%%%%%%%%%%%%%%%%%%%%%%%%%%%%%%%%%%%

\subsection{Color spread and Mg abundance for bright RGB stars}

%%%%%%%%%%%%%%%%%%%%%%%%%%%%%%%%%%%%%%%%%%%%%%%%%% FIG 
\begin{figure}
\includegraphics[width=80mm]{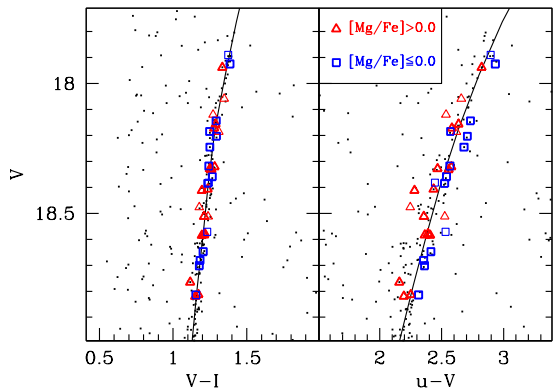}
 \caption{Location of Mg-deficient (blue squares) and Mg-rich (red-triangles) stars from \citet{muccia} in the V, V-I (left panel) and V, u-V (right panel) CMDs; the solid lines are the RGB ridge lines. Lighter symbols correspond to stars that are present in our catalog but are excluded from the analyzed sample because of the selections in {\em CHI} and {\em SHARP} adopted in Sect.~\ref{sel}. 
 }
 \label{cmdmag}
\end{figure}
%%%%%%%%%%%%%%%%%%%%%%%%%%%%%%%%%%%%%%%%%%%%%%%%%%%%%%%%%%%%%%%%%%%%%%%

In fig.~\ref{cmdmag} we show the location in the V, V-I and V, u-V CMDs of the stars studied by Mu12, according to their Mg abundance. We limited our comparison to stars fainter than $V=17.85$
to avoid spurious effect tied to the partial saturation affecting stars brighter than this limit (see Sect.~\ref{cmd}). For the same reason, we cannot perform any useful test of correlation between chemical abundances and color for the \citet{judy12} sample, since only 3 of the 13 stars in the sample are fainter than $V=17.85$. 

It should be noted that, in spite of the magnitude cut we adopted, several of the Mu12 stars shown in Fig.~\ref{cmdmag} do not pass the quality criteria described in Sect.~\ref{sel}. Still it is hard to conceive that the color segregation between Mg-deficient and Mg-normal\footnote{Adopting the nomenclature by Mu12.} stars emerging in the V, u-V CMD (and not seen in the V, V-I one) of Fig.~\ref{cmdmag} can be a spurious effect of larger uncertainties. In the V, u-V CMD 13 of the 15 Mg-deficient stars lie to red of the ridge line, while 
13 of 19 Mg-normal stars lie to the blue of the ridge line. We recall that the [Mg/Fe] distribution in NGC~2419 is bi-modal, with a threshold between the two groups occurring at [Mg/Fe]$\simeq 0.0$: hence the separation adopted in Fig.~\ref{cmdmag} reflects a physical separation between two generations of stars.

In the typical GC self-enriched in light elements (and not in iron), Mg-poor stars would be also Na-rich, hence the results shown in Fig.~\ref{cmdmag} would be strictly analogous to those shown in other clusters, where Na-rich RGB stars have redder near-UV-optical color than Na-poor ones at the same magnitude \citep[see, e.g.,][L11]{fabi_m4}. 

However, with very few exceptions, in typical GCs Mg-poor/Na-rich stars have [Mg/Fe]$>0.0$, the large population of Mg-deficient ([Mg/Fe]$<0.0$) stars is a unique characteristic of NGC~2419. Moreover, while the small sample studied by \citet{judy12} does not allow to draw firm conclusions, no correlation between Mg and Na abundances emerges from their data \citep[see][]{ventura}. 
Finally, the results presented in the previous sections seem to confirm the notion that the main factor driving the color spread in the RGB in NGC~2419 is the spread in He abundance, not in CN and NH bands strength, as in other clusters. Therefore, it is unlikely that the general scheme for the interpretation of the behavior of typical clusters can be applied in the present case. 
It has also to be noted that the stars considered in this section lie in a completely different magnitude range with respect to those considered in previous sections ($17.85\le V\le 18.85$ vs. $19.8\le V\le 22.0$). By analogy with the results of Mi12, it is quite likely that the two effects are due to different factors acting in different regions of the CMDs.

To investigate the origin of the correlation between Mg abundance and u-V color on the upper RGB shown in Fig.~\ref{cmdmag} one should try to couple stellar interior and  atmosphere models computed on purpose to reproduce the unique (and puzzling) chemical composition of this cluster \citep[Mu12,][]{judy12}, that is clearly beyond the scope of the present analysis.

\section{Summary and Conclusions}

Using a large number of deep multi-band images from LBC on LBT, we obtained accurate u, V and I
photometry of the Red Giant Branch of the massive globular cluster NGC~2419. The main scientific goal of our observations was to search for near-UV color spreads at fixed magnitude on the RGB, a typical photometric signature of multiple populations in GCs (L11). We found a very significant color spread, in excess of what expected from observational errors, in all the considered colors, u-I, u-V, V-I. Our findings are in qualitative and quantitative agreement with the results by D11, who proposed that the RGB color spread in this cluster is mainly due to a large spread of Helium abundance between the subsequent stellar generations of the cluster. The occurrence of a significant spread also in colors not including near-UV filters, found by D11 with various passbands combination and confirmed here in V-I, strongly support this view. The variations of NH and CN band strength that drive the near-UV color spread in typical clusters is ineffective in optical colors (see, e.g., L11) , while variations in He change the surface temperature of the stars independently of the details of the stellar atmosphere \citep[see][for discussion and references]{santi}.
In this scenario, the He-rich second-generation RGB stars lie to the blue of the He-normal RGB stars from the first generation, as shown in Fig.~\ref{cmdhe}, above.

We divided our RGB sample into a RGB-Blue and a RGB-Red subsample, according to their position in (u-V or u-I) color with respect to the cluster ridge line, and we investigated their radial distributions. We found that these differ at a very high level of significance, RGB-Blue stars being more centrally concentrated than RGB-Red ones. {\em This is the first detection of a difference in the radial distribution between different samples of stars in this cluster}. Interpreting the RGB-Blue sample as dominated by second generation stars, following D11, the sense of the observed difference is the same found in all the other GCs where differences in radial distributions have been detected, i.e. {\em second generation stars are more centrally concentrated than first generation stars}, in good agreement with the recent models by \citet{vespe}. It has to be noted that the difference in the radial distribution of first and second generation stars at the end of the star formation epoch, is a general prediction of all the models of chemical self-enrichment in GCs \citep{FRMS,decres08,AGB,selma}

The lack of a spread in iron abundance combined with the significant spread in He and light elements, and the difference in the radial distribution between RGB-Blue and RGB-Red stars strongly suggest that the peculiar abundance pattern observed in NGC~2419 is more likely due to an extreme manifestation of the multiple population syndrome affecting GCs than to the self-enrichment processes at work in typical dwarf galaxies \citep[D11,Mu12][]{judy12,ventura}. 
The interpretative scheme recently proposed by \citet{ventura} can possibly accommodate also the small spread in Calcium abundance claimed by \citet{judy12}\footnote{The only difference in the results 
of the spectroscopic analyses by Mu12 and \citet{judy12}) is about the spread in the Ca abundance: the former conclude that Ò... the spread in calcium abundance is absent or very small...Ó, while \citet{judy12} claims that a small but real spread is actually there, especially among Mg-poor stars. From their data, using the Maximum Likelihood technique adopted by Mu12, we find that the intrinsic Ca spread is $\sigma[Ca/F e] = 0.05 \pm 0.13$.} into a chemical enrichment path driven by AGB stars. 

However it should be recalled that, while promising, Ventura et al.'s model is highly speculative
(for instance cross sections of relevant nuclear reaction must be stretched by a factor of a hundred with respect to their standard value, to reproduce the observed enhancement in Potassium abundance)
and we are far from a complete understanding of the evolutionary path of NGC~2419. It is quite likely that, in spite of the impressive progresses of the latest years, we are also still lacking important observational facts, as emphasized by \citet{ventura}. The correlation between u-V color and Mg abundance presented in Fig.~\ref{cmdmag} provide an excellent example in this sense. While the effect is analogous to the correlation between near-UV colors and, e.g., Na abundance observed in many metal-intermediate and metal-rich clusters, it is very unlikely that it has the same origin.
Stellar models predict that the He-rich RGBs from the second generation should lie to the red of first generation stars all along the RGB (see Fig.~\ref{cmdhe}). Hence Mg-poor stars, that should belong to second generation \citep{ventura}, should lie on the blue side of the RGB, since the effects of NH and CN bands should be negligible at the metallicity of NGC~2419. Still we observe that  Mg-poor stars lie in the red side of the RGB, indicating that some other mechanisms should be at work, possibly some unexpected spectral feature due to the the highly non-standard composition of the atmosphere of the stars of this cluster. Obtaining spectra of these stars in the region of the U band seems the only way to get further insight on this puzzling issue.

\section*{Acknowledgments}

We acknowledge the financial support of INAF through the PRIN-INAF
2009 grant assigned to the project ``Formation and evolution of massive star
clusters'', P.I.: R. Gratton. GB acknowledges the European CommunityÕs Seventh Framework 
Programme under grant agreement no.229517. 
EC acknowledges the PRIN INAF 2011 ``Multiple populations in globular
clusters: their role in the Galaxy assembly'' (PI E. Carretta);
MB, AB, EC acknowledge the PRIN MIUR 2010-2011, ``The Chemical
and Dynamical Evolution of the Milky Way and Local Group Galaxies'',
(PI F. Matteucci), prot. 2010LY5N2T.
This research has made use of NASA's Astrophysics Data System.

\label{lastpage}

\end{document}